\documentclass[11pt,a4paper,english,nofootinbib,,superscriptaddress]{revtex4}
\usepackage{lmodern}

\usepackage[T1]{fontenc}
\usepackage[latin9]{inputenc}
\usepackage{babel}
\usepackage{slashed}
\usepackage{amsmath}
\usepackage{graphicx}
\usepackage{amssymb}
\usepackage{esint}
\usepackage[unicode=true, pdfusetitle,
 bookmarks=true,bookmarksnumbered=false,bookmarksopen=false,
 breaklinks=false,pdfborder={0 0 1},backref=false,colorlinks=false]
 {hyperref}
\def\b{\begin{equation}}
\def\e{\end{equation}}

\makeatletter
\@ifundefined{textcolor}{}
{%
 \definecolor{BLACK}{gray}{0}
 \definecolor{WHITE}{gray}{1}
 \definecolor{RED}{rgb}{1,0,0}
 \definecolor{GREEN}{rgb}{0,1,0}
 \definecolor{BLUE}{rgb}{0,0,1}
 \definecolor{CYAN}{cmyk}{1,0,0,0}
 \definecolor{MAGENTA}{cmyk}{0,1,0,0}
 \definecolor{YELLOW}{cmyk}{0,0,1,0}
 }

\usepackage{latexsym}\usepackage{bm}

\makeatother

\makeatother

\begin{document}
	
\begin{flushright}{
		MIT-CTP-4768\\}
	
\end{flushright}	

\title{Faddeev-Jackiw Hamiltonian Reduction for Free and Gauged Rarita-Schwinger Theories}

\author{Suat Dengiz}

\email{sdengiz@mit.edu}

\affiliation{Center for Theoretical Physics,\\
	 Massachusetts Institute of Technology, Cambridge MA 02139 USA}

\date{\today}
\begin{abstract}
We study the Faddeev-Jackiw symplectic Hamiltonian reduction for $3+1$-dimensional free and Abelian gauged Rarita-Schwinger theories that comprise Grassmannian fermionic fields.
We obtain the relevant fundamental brackets and find that they are in convenient forms for quantization. The brackets are independent of whether the theories contain mass or gauge fields, and the structure of constraints and symplectic potentials largely determine characteristic behaviors of the theories. We also note that, in contrast to the free massive theory, the Dirac field equations for free \emph{massless} Rarita-Schwinger theory cannot be obtained in a covariant way.  
\end{abstract}
\maketitle

\section{Introduction}

In 1941, Rarita and Schwinger constructed a theory of spin-$\frac{3}{2}$ vector-spinor fields which has a local fermionic gauge-invariance \cite{RaritaSchwinger}. However, this symmetry is lost when the vector-spinor field has mass or couples to the other lower spin fields. More precisely, in 1961, Johnson and Sudarshan studied massive Rarita-Schwinger field minimally coupled to an external electromagnetic field, and showed that the equal-time commutators and relativistic covariance of the theory are in conflict, which makes the quantization a rather subtle issue \cite{JohnsonSudarshan}. In 1969, Velo and Zwanziger found that the massive gauged extension of the theory also admits superluminal wave propagation. Thus, the causality principle is also violated in the theory \cite{VeloZwanziger}. Despite these persistent problems, the massless theory keeps its importance particularly in two aspects. First, the massless (Majorana) Rarita-Schwinger field plays a central role in the construction of covariantly interacting supergravity theory \cite{FreedmanNieu, GrisaruPendletonNieuwenhuizen, GrisaruPendleton}. The theory describes a generalization of the Rarita-Schwinger fermionic gauge-invariance and the vector-spinor fields are fermionic superpartner of gravitons, namely gravitinos of the supergravity. In this concept, Das and Freedman showed that the massless theory is free from the non-causal wave propagation and has a unitary propagator structure \cite{ DasFreedman}. Secondly, the massless Rarita-Schwinger theory is valuable for the cancellation of SU(8) gauge anomalies. Unlike the generic anomaly cancellation mechanisms in which the anomalies are supposed to be canceled withing the lower spin fermionic fields, it was shown by Marcus \cite{Marcus} and later studied by Adler \cite{Adler2014}, that a complete SU(8) gauge theory can be constructed via Rarita-Schwinger fields. In this set-up, the vector-spinor field acquires a crucial role in canceling anomalies arising in the gauge theory. Thus, it is left to determine whether the gauged Rarita-Schwinger fields describe well-behaved, complete classical or quantum field theories. For this purpose, Adler has recently studied minimally gauged massless Rarita-Schwinger theories at both classical and quantum levels in detail \cite{Adler2015}. He showed that, unlike the massive case, the massless gauged Rarita-Schwinger theory provides consistent classical and quantum theories with a generalized fermionic gauge-invariance.

Taking the above mentioned observations as inspiration points and noting the hard task of getting proper brackets of constrained systems providing viable quantization, we study the Faddeev-Jackiw (FJ) symplectic Hamiltonian reduction \cite{FaddeevJackiw, Jackiwbook} for free and gauged Rarita-Schwinger theories. Unlike Dirac's approach for constrained systems \cite{Dirac}, FJ symplectic first-order formalism does not require any classification of constraints\footnote{For the quantization of the constrained system, see for example \cite{Henneaux, Fradkin, Batalin}.}. In other words, the method avoids analyzing systems by evaluating all commutation relations among the constraints and classifying them accordingly. Apparently, the FJ approach supplies a rather economical way of quantizing constrained systems. In doing so, we find the fundamental brackets for the free and gauged Rarita-Schwinger theories for both massless and massive versions. Here, the brackets are in admissible structures to be quantized. We also observe that the brackets are identical for all kinds of the theories; the brackets are independent of whether the theory is massive or interacting with external electromagnetic field or not. The differences between the theories arise among the constraints they have. We also notice that, in contrast to the massive case, the Dirac field equations for free \emph{massless} Rarita-Schwinger theory cannot be obtained in a covariant way.

The layout of the paper is as follows: In Sec. II, we recapitulate the fundamental properties of free massless Rarita-Schwinger theory and apply FJ Hamiltonian reduction to the theory. In Sec. III, we turn our attention to the FJ Hamiltonian reduction for free massive Rarita-Schwinger theory. Sec. IV and Sec. V are devoted to the first-order symplectic analysis for Abelian gauged extensions of massless and massive Rarita-Schwinger theories. In Sec. VI, we conclude our results. In the Appendix A, the derivation of the transverse and traceless decomposition of the fields in the free massless Rarita-Schwinger theory is given as a sample. In the Appendix B, we briefly review the FJ approach for constrained and unconstrained systems. We also give an example of the application of symplectic method to anti-commuting spin-$\frac{1}{2}$ Dirac theory.

\section{Free Massless Rarita-Schwinger Theory}

The $3+1$-dimensional free massless Rarita-Schwinger theory is described by the Lagrangian 
\begin{equation}
{\cal L}=-\epsilon^{\lambda\mu\nu\rho} \bar{\psi}_\lambda \gamma_5 \gamma_\mu \partial_\nu \psi_\rho,
\label{freemasslesrst}
\end{equation}
where $\psi_\mu $ and $\bar{\psi}_\mu$ are vector-spinor fields with spinor indices suppressed. We work in the metric signature $(+, -, -, -)$, $  \gamma_5=\rm{i} \gamma^0 \gamma^1 \gamma^2 \gamma^3$ and $ \{\gamma^\mu, \gamma^\nu \}=2 \eta^{\mu\nu} $. We consider the fermionic fields as independent anti-commuting Grassmannian variables. Recall that, unlike the complex Dirac field, for the Grassmannian variables there is no such relation as $\bar{\psi}_\mu=\gamma^0 \psi^+_\mu $. Instead, $ \psi_\mu $ and $\bar{\psi}_\mu $ are independent generators in the Grassmann algebra. Thus, one can define the \emph{conjugation} as follows:
\begin{equation}
\psi^*_\mu=\bar{\psi}_\nu (\gamma^0)^\nu{_\mu}, \hskip 1 cm  (\bar{\psi}_\mu)^*=(\gamma^0)_\mu{^\nu} \psi_\nu. 
\label{conjugat}
\end{equation}
Notice that this does \emph{not} mean that Eq.(\ref{conjugat}) produces a new element in the Grassmannian algebra. This is merely the conjugation of independent variables. Therefore, with the help of the conjugation of the Grassmannian variables $(\theta_1 \theta_2)^*=\theta^*_2 \theta^*_1 $, one can show that the Lagrangian in Eq.(\ref{freemasslesrst}) is self-adjoint up to a boundary term:
\begin{equation}
{\cal L}^*={\cal L}+ \partial_\nu (\epsilon^{\lambda\mu\nu\rho} \bar{\psi}_\lambda \gamma_5 \gamma_\mu  \psi_\rho ),
\end{equation}
such that the total derivative term naturally drops at the action level. Moreover, variations with respect to independent variables respectively yield
\begin{equation}
	\epsilon^{\lambda\mu\nu\rho} \gamma_5 \gamma_\mu \partial_\nu \psi_\rho =0, \qquad \epsilon^{\lambda\mu\nu\rho} \partial_\nu \bar{\psi}_\lambda \gamma_5 \gamma_\mu =0,
	\label{fieldeqfrst1}
\end{equation}
which are the corresponding field equations. From now on, we will work with the first of Eq.(\ref{fieldeqfrst1}). But, by following the same steps, one could easily obtain the similar results for the second equation. Notice that by using the identity
\begin{equation}
\epsilon^{\lambda\mu\nu\rho} \gamma_5 \gamma_\mu=\rm{i} ( \eta^{\lambda\rho}\gamma^\nu-\eta^{\lambda\nu}\gamma^\rho-\gamma^\lambda \eta^{\rho\nu}+\gamma^\lambda \gamma^\nu \gamma^\rho ),
\label{identity}
\end{equation}
one can recast the field equation in Eq.(\ref{fieldeqfrst1}) as follows 
\begin{equation}
\slashed{\partial} \psi^\lambda- \partial^\lambda (\gamma \cdot \psi )-\gamma^\lambda \partial \cdot \psi +\gamma^\lambda \slashed{\partial} (\gamma \cdot \psi)=0.
\label{fieldeqfrst2}
\end{equation}
Here $ \slashed{\partial} =\gamma^\mu \partial_\mu $ and $ \gamma \cdot \psi =\gamma^\mu \psi_\mu $. Contracting Eq.(\ref{fieldeqfrst2}) with $\gamma_\lambda$ gives
\begin{equation}
\partial \cdot \psi -\slashed{\partial} (\gamma \cdot \psi)=0.
\label{eqncontrgammalambd}
\end{equation}
Finally, by plugging this result in Eq.(\ref{fieldeqfrst2}), the field equation reduces to
\begin{equation}
\slashed{\partial} \psi^\lambda-\partial^\lambda (\gamma \cdot \psi)=0.
\label{ferswthcons}
\end{equation}
To obtain the real propagating degrees of freedom, let us now study gauge transformation and corresponding gauge conditions. For this purpose, let us recall that under the local Rarita-Schwinger fermionic gauge transformation
\begin{equation}
\delta \psi_\rho(x) =\partial_\rho \epsilon (x),
\label{gaugtrffreemasslesrst}
\end{equation}
the Lagrangian in Eq.(\ref{freemasslesrst}) transforms as
\begin{equation}
\delta {\cal L}=\partial_\lambda (- \epsilon^{\lambda\mu\nu\rho} \bar{\epsilon} \gamma_5 \gamma_\mu \partial_\nu \psi_\rho).
\label{rargagtransf}
\end{equation}
Here $\epsilon$ is an arbitrary four-component spinor field. As is seen in Eq.(\ref{rargagtransf}), the free massless Rarita-Schwinger Lagrangian changes by a total derivative under the Rarita-Schwinger gauge transformation, which drops at the action level and thus we have a completely gauge-invariant theory. This means that the theory admits a gauge redundancy.
To find the correct physical degrees of freedom of the theory, one needs to fix this gauge-freedom. For this purpose, let us assume the Coulomb-like gauge condition
\begin{equation}
\gamma^i \psi_i=0,
\label{gaugfixfrst}
\end{equation}
where $i=1,2,3$. In fact, this is a reasonable gauge choice: Any initial data $\psi^{'}_i({\bf x}, t)$ that does not satisfy Eq.(\ref{gaugfixfrst}) can be tuned to the desired form via \footnote{Since the gauge choice $\partial^i \psi_i=0$ on the initial data will also arise due to the self-consistency, one should also be able to regulate the gauge parameter via $ \epsilon=-\frac{1}{\nabla^2} \partial_i \psi^i $. But since we start with the (\ref{gaugfixfrst}), we have to give (\ref{gagparreg}).}
\begin{equation}
\epsilon({\bf x}, t)=-\gamma^i \partial_i \int \frac{d^3 y}{4\pi \lvert{\bf x} -{\bf y}\rvert} \gamma^j \psi_j({\bf y}, t). 
\label{gagparreg}
\end{equation}
(See \cite{DasFreedman} and \cite{FreedmanVanProyen} for further discussions). For the sake of the self-completeness, one needs to examine the theory further to see whether Eq.(\ref{gaugfixfrst}) imposes any additional conditions or not. For this purpose, note that $\psi_0$ component does not have a time derivative, so it is a Lagrange multiplier. In other words, as in the electromagnetic case, the zeroth component of the vector-spinor field is a zero mode which is followed with a constraint. More precisely, the $\lambda=0$ component of the field equation in Eq.(\ref{ferswthcons}) reads 
\begin{equation}
\gamma^i \partial_i \psi_0-\partial_0 (\gamma^i \psi_i) =0.
\label{zerothcomp1}
\end{equation}
One can also get a secondary constraint by contracting the field equation with $\partial_\lambda $. But since our primary aim is not analyzing the system by examining all the existing constraints, we leave it as a comment. As is seen in Eq.(\ref{zerothcomp1}), gauge fixing condition $\gamma^i \psi_i=0 $ imposes $ \gamma^i \partial_i \psi_0=0 $. Here, since the operator is not invertible, we are not allowed to get $\psi_0=0$ as a corollary of $\gamma^i \psi_i=0 $; yet we assume an additional condition of $\psi_0=0$. Furthermore, splitting the fully contracted equation in Eq.(\ref{eqncontrgammalambd}) into its space and time components yields
\begin{equation}
\partial^i \psi_i-\gamma^0 \partial_0 (\gamma^i \psi_i)-\gamma^i \partial_i (\gamma^0 \psi_0)-\gamma^i \partial_i (\gamma^j \psi_j)=0. 
\label{timspdecfulcontr}
\end{equation}
In Eq.(\ref{timspdecfulcontr}), one should notice that the gauge fixing condition $\gamma^i \psi_i=0$ together with the assumed condition $ \psi_0 =0$ impose $ \partial^i \psi_i=0 $. As a consequence of this, we obtain the set of consistency conditions 
\begin{equation}
\gamma^i \psi_i=0 \quad , \quad \partial^i \psi_i=0 \quad , \quad  \psi_0=0.
\label{conssconds}
\end{equation}
Observe that Eq.(\ref{conssconds}) can also be written in covariant forms as follows
\begin{equation}
\gamma^\mu \psi_\mu=0 \quad , \quad \partial^\mu \psi_\mu=0,
\label{covconscond}
\end{equation}
which are the Rarita-Schwinger gauge fixing conditions. Thus, with the gauge choices in Eq.(\ref{covconscond}), the field equation for the free massless Rarita-Schwinger theory in Eq.(\ref{ferswthcons}) turns into the Dirac field equation for massless spin-$\frac{3}{2}$ vector-spinor field
\begin{equation}
	\slashed{\partial} \psi^\lambda = 0.
\end{equation}

\subsection*{Symplectic Reduction for Free Massless Rarita-Schwinger Theory}   

In this section, we study the FJ Hamiltonian reduction for the free massless Rarita-Schwinger theory which will lead us to the fundamental brackets of the theory. For this purpose, let us recast the Lagrangian in Eq.(\ref{freemasslesrst}) in a more symmetric form:
\begin{equation}
{\cal L}=-\frac{1}{2} \epsilon^{\lambda\mu\nu\rho} \bar{\psi}_\lambda \gamma_5 \gamma_\mu \partial_\nu \psi_\rho + \frac{1}{2} \epsilon^{\lambda \mu \nu\rho} (\partial_\nu \bar{\psi}_\lambda) \gamma_5 \gamma_\mu \psi_\rho.
\label{fjfrmasllsrst1}
\end{equation}
To study the theory in the first-order symplectic formalism, one needs to convert Eq.(\ref{fjfrmasllsrst1})  into the desired symplectic form. That is, one needs to split the Lagrangian into its space and time components. After a straightforward decomposition, one gets 
\begin{equation}
{\cal L}={\cal A}^{(k)}_1 \dot{\psi}_k + {\cal A}^{(k)}_2\dot{\bar{\psi}}_k-{\cal H}(\psi_0, \bar{\psi}_0, \psi_k, \bar{\psi}_k),
\label{timspacdecfremsslsssrs123}
\end{equation}
where the symplectic coefficients are
\begin{equation}
{\cal A}^{(k)}_1=-\frac{1}{2} \epsilon^{ijk} \bar{\psi}_i \gamma_5 \gamma_j, \qquad {\cal A}^{(k)}_2=\frac{1}{2} \epsilon^{ijk}\gamma_5 \gamma_j \psi_i,
\end{equation}
and the corresponding symplectic potential reads 
\begin{equation}
\begin{aligned}
{\cal H}(\psi_0, \bar{\psi}_0, \psi_k, \bar{\psi}_k)&=\frac{1}{2} \epsilon^{ijk} \bar{\psi}_0 \gamma_5 \gamma_i \partial_j \psi_k-\frac{1}{2} \epsilon^{ijk} \bar{\psi}_i \gamma_5 \gamma_0 \partial_j \psi_k-\frac{1}{2} \epsilon^{ijk} \bar{\psi}_i \gamma_5 \gamma_j \partial_k \psi_0 \\
&-\frac{1}{2} \epsilon^{ijk} (\partial_j \bar{\psi}_0) \gamma_5 \gamma_i  \psi_k+\frac{1}{2} \epsilon^{ijk} (\partial_j \bar{\psi}_i) \gamma_5 \gamma_0 \psi_k+\frac{1}{2} \epsilon^{ijk} (\partial_k \bar{\psi}_i) \gamma_5 \gamma_j  \psi_0.
\end{aligned}
\end{equation}
As expected, all the non-dynamical components have been relegated into the Hamiltonian part of the system. In analyzing the theory, one could also choose the conjugate momenta of $\bar{\psi}_k$ as a dynamical variable. But in our analysis, we will not work with it. Instead, we consider $\psi_\mu$ and $\bar{\psi}_\mu $ as the independent variables. Note that $\psi_0$ and $\bar{\psi}_0$ are not dynamical components, so they are Lagrange multipliers. Following \cite{FaddeevJackiw, Jackiwbook}, the elimination of constraints give the equations
\begin{equation}
\epsilon^{ijk}(\partial_k \bar{\psi}_i) \gamma_5 \gamma_j =0, \qquad \quad \epsilon^{ijk} \gamma_5 \gamma_i \partial_j \psi_k =0.
\label{constrfreemasslessrrs12}
\end{equation}
To solve the constraint equations, one can decompose the independent fields into its \emph{local transverse and $\gamma$-traceless parts} as 
\begin{equation}
\psi_i =\psi^T_i+\hat{\psi}_i \qquad \bar{\psi}_i=\bar{\psi}^T_i+\hat{\bar{\psi}}_i,
\label{ttfreemassless12}
\end{equation}	
where "$T$" and "$\textasciicircum$" stand for the transverse and traceless parts, respectively. Here the $\gamma$-traceless parts are
\begin{equation}
\hat{\psi}_i=\psi_i-\frac{1}{3} \gamma_i \gamma^j \psi_j, \hskip 1 cm \hat{\bar{\psi}}_i=\bar{\psi}_i-\frac{1}{3} \bar{\psi}_j \gamma^j \gamma_i,
\end{equation}	
such that $ \gamma^i \hat{\psi}_i=0 $ and $\gamma^i \hat{\bar{\psi}}_i=0  $. Then, by using the identity
\begin{equation}
\epsilon^{ijk} \gamma_5 \gamma_k=-\gamma^0 \sigma^{ij} \quad  \mbox{where} \quad \sigma^{ij}=\frac{i}{2} [\gamma^i, \gamma^j],
\label{identity1}
\end{equation}
as well as the constraints in Eq.(\ref{constrfreemasslessrrs12}), one can show that the transverse and traceless decomposition of the fields in Eq.(\ref{ttfreemassless12}) can actually be written as follows 
\begin{equation}
\psi_i = \psi^T_i+\frac{\partial_i \zeta}{\nabla^2}, \qquad \quad \bar{\psi}_i = \bar{\psi}^T_i+\frac{\partial_i \bar{\zeta}}{\nabla^2},
\label{ttdecfrersmsslssd1}
\end{equation}
where $\zeta=\slashed{\partial}(\gamma \cdot \psi^T) $ and $\nabla^2=\partial_i \partial^i$. As a side comment, one should note that as is done in \cite{Jackiwbook}, without addressing the transverse and $\gamma$-traceless parts (\ref{ttfreemassless12}), one could also directly start with the (\ref{ttdecfrersmsslssd1}). Here, we further provide what the explicit form of the Longitudinal part is. (See Appendix A for the derivation of Eq.(\ref{ttdecfrersmsslssd1})). Accordingly, the constraint equations in Eq.(\ref{constrfreemasslessrrs12}) turn into completely transverse ones
\begin{equation}
\epsilon^{ijk}(\partial_k\bar{\psi}^T_i) \gamma_5 \gamma_j=0, \hskip 1 cm \epsilon^{ijk} \gamma_5 \gamma_i \partial_j \psi^T_k=0.
\label{constrfreemasslessrrs22}
\end{equation}

Finally, by inserting Eq.(\ref{ttdecfrersmsslssd1}) and Eq.(\ref{constrfreemasslessrrs22}) in the Eq.(\ref{timspacdecfremsslsssrs123}), up to a boundary term, one gets a completely transverse Lagrangian
\begin{equation}
{\cal L}={\cal A}^{(k)^T}_1 \dot{\psi}^T_k + {\cal A}^{(k)^T}_2 \dot{\bar{\psi}}^T_k-{\cal H}_T(\psi^T_k, \bar{\psi}^T_k).
\end{equation}
Here the transverse symplectic coefficients and potential are
\begin{equation}
\begin{aligned}
{\cal A}^{(k)^T}_1&=-\frac{1}{2} \epsilon^{ijk} \bar{\psi}^T_i \gamma_5 \gamma_j, \hskip 1.2 cm  {\cal A}^{(k)^T}_2=\frac{1}{2} \epsilon^{ijk}\gamma_5 \gamma_j \psi^T_i, \\
{\cal H}_T(\psi^T_k, \bar{\psi}^T_k) & =-\frac{1}{2} \epsilon^{ijk} \bar{\psi}^T_i \gamma_5 \gamma_0 \partial_j \psi^T_k+\frac{1}{2} \epsilon^{ijk} (\partial_j \bar{\psi}^T_i) \gamma_5 \gamma_0 \psi^T_k.
\end{aligned}
\end{equation}
Thus, by defining the symplectic variables as $ (\xi_1, \xi_2)=(\psi^T_k, \bar{\psi}^T_k) $, one gets the corresponding symplectic matrix 
\[ f_{\alpha\beta}= \left( \begin{array}{cc}
0  & \epsilon^{ijk} \gamma_5 \gamma_j  \\
-\epsilon^{ijk} \gamma_5 \gamma_j & 0 \end{array} \right)=\epsilon_{\alpha\beta}\epsilon^{ijk} \gamma_5 \gamma_j ,\]
which is clearly non-singular. Notice that the minus sign in the sub-block is due to the anti-symmetric $\epsilon$ tensor. Therefore, by taking care of the epsilons contraction in the current signature, one can easily show that the inverse symplectic matrix is 
\[ f^{-1}_{\alpha\beta }= \left( \begin{array}{cc}
0  & -\frac{1}{2}\epsilon_{imk} \gamma_5 \gamma^m  \\
\frac{1}{2}\epsilon_{imk} \gamma_5 \gamma^m  & 0 \end{array} \right)=\frac{1}{2}\epsilon_{\beta\alpha}\epsilon_{imk} \gamma_5 \gamma^m.  \]
Once the inverse symplectic matrix is found, one can evaluate the fundamental brackets. That is, by using the definition of the FJ equal-time brackets for the Grassmann variables 
\begin{equation}
\{\xi_\beta, \xi_\alpha\}_{FJ}=-f^{-1}_{\alpha\beta},
\end{equation}
one gets the fundamental brackets for free massless Rarita-Schwinger theory as follows
\begin{equation}
\begin{aligned}
& \{\psi^T_i(x),\bar{\psi}^T_k(y) \}_{FJ}=-\frac{1}{2}\epsilon_{imk} \gamma_5 \gamma^m  \delta^3(x-y), \\
& \{\psi^T_i(x),\psi^T_k(y) \}_{FJ}=0, \hskip 1 cm \{\bar{\psi}^T_i(x),\bar{\psi}^T_k(y) \}_{FJ}=0.
\end{aligned}
\end{equation}
Note that, with the help of the identity in Eq.(\ref{identity1}), the non-vanishing bracket can also be rewritten as 
\begin{equation}
\{\psi^T_i(x),\bar{\psi}^T_k(y) \}_{FJ}=\frac{\rm{i}}{2} \gamma_k \gamma_i \gamma_0 \delta^3(x-y),
\label{fjbracfreemsslssrs}
\end{equation} 
which is identical with the one found in \cite{Okawa}.

\section{Free Massive Rarita-Schwinger Theory} 

The Lagrangian that describes the $3+1$-dimensional free massive Rarita-Schwinger theory is 
\begin{equation}
	{\cal L}=-\epsilon^{\lambda\mu\nu\rho}\bar{\psi}_\lambda \gamma_5 \gamma_\mu \partial_\nu \psi_\rho+\rm{i} \mbox{m} \bar{\psi}_\lambda \sigma^{\lambda\rho} \psi_\rho,  
	\label{lagfrmassvrs}
\end{equation}
where $\sigma^{\lambda\rho}=\frac{\rm{i}}{2} [\gamma^\lambda, \gamma^\rho]=\rm{i}(\eta^{\lambda \rho}-\gamma^\rho \gamma^\lambda) $. Recall that the fermionic fields are anti-commuting Grassmannian variables. Accordingly, the field equations of the independent variables respectively read
\begin{equation}
	\epsilon^{\lambda\mu\nu\rho} \gamma_5 \gamma_\mu \partial_\nu \psi_\rho-\rm{i} m \sigma^{\lambda\rho} \psi_\rho =0, \hskip 1cm  
	\epsilon^{\lambda\mu\nu\rho} \partial_\nu \bar{\psi}_\lambda \gamma_5 \gamma_\mu  + \rm{i} m \bar{\psi}_\lambda \sigma^{\lambda\rho} =0.
	\label{freemassfeqfirst}
\end{equation}  
In dealing with the fundamental properties of the theory, as we did in the massless theory, we will work only with the first field equation in Eq.(\ref{freemassfeqfirst}). Notice that by using the identity in Eq.(\ref{identity}), one can recast the field equation as follows
\begin{equation}
	\rm{i} [\slashed{\partial} \psi^\lambda- \partial^\lambda (\gamma \cdot \psi )-\gamma^\lambda \partial \cdot \psi +\gamma^\lambda \slashed{\partial} (\gamma \cdot \psi) ]-\rm{i} m \sigma^{\lambda\rho} \psi_\rho =0.
	\label{fieldeqfrst22}
\end{equation}
Observe that the contraction of Eq.(\ref{fieldeqfrst22}) with $ \gamma_\lambda $ yields
\begin{equation}
2\rm{i} [\slashed{\partial}(\gamma \cdot \psi)-\partial \cdot \psi ]+3 \rm{m} \gamma \cdot \psi =0,
\label{contemomassive1}
\end{equation}
and the contraction of Eq.(\ref{fieldeqfrst22}) with $ \partial_\lambda $ gives
\begin{equation}
\rm{m} [\slashed{\partial}(\gamma \cdot \psi)-\partial \cdot \psi ]=0.
\label{contemomassive2}
\end{equation}
Combining both contracted field equations Eq.(\ref{contemomassive1}) and Eq.(\ref{contemomassive2}), one obtains 
\begin{equation}
	\gamma \cdot \psi =0 ,\qquad \partial \cdot \psi=0.
\end{equation}
With these gauge-fixing conditions, the equation in Eq.(\ref{fieldeqfrst22}) turns into the Dirac field equation for massive spin-$\frac{3}{2}$ vector-spinor field  
\begin{equation}
	(\rm{i} \slashed{\partial} +m )\psi^\lambda=0.
	\label{dirlikmassvfrrs}
\end{equation} 
Note that, unlike the massless theory, one obtains the Dirac field equation in Eq.(\ref{dirlikmassvfrrs}) without addressing the space and time decompositions of the field equations. On the other hand, due to the mass term, the Rarita-Schwinger gauge-invariance is inevitably lost.

\subsection*{Symplectic Reduction for Free Massive Rarita-Schwinger Lagrangian}

Let us now study the symplectic Hamiltonian reduction of the free massive Rarita-Schwinger theory. For this purpose, let us recall that the Lagrangian in Eq.(\ref{lagfrmassvrs}), up to a boundary term, can be written as 
\begin{equation}
{\cal L}=-\frac{1}{2}\epsilon^{\lambda\mu\nu\rho}\bar{\psi}_\lambda \gamma_5 \gamma_\mu \partial_\nu \psi_\rho+\frac{1}{2}\epsilon^{\lambda\mu\nu\rho} \partial_\nu \bar{\psi}_\lambda \gamma_5 \gamma_\mu \psi_\rho+\rm{i} \rm{m} \bar{\psi}_\lambda \sigma^{\lambda\rho} \psi_\rho.
\label{fjfrmassivesrst1}
\end{equation}
In order to proceed the FJ symplectic reduction of Eq.(\ref{fjfrmassivesrst1}), one needs to separate the dynamical components from the non-dynamical ones so that the non-dynamical components can be relegated to Hamiltonian part of the Lagrangian. Therefore, by splitting the Lagrangian into its space and time components, one will obtain
\begin{equation}
{\cal L}={\cal A}^{(k)}_1 \dot{\psi}_k + {\cal A}^{(k)}_2 \dot{\bar{\psi}}_k-{\cal H}(\psi_0, \bar{\psi}_0, \psi_k, \bar{\psi}_k),
\label{massivelagrangian32}
\end{equation}
where the coefficient of the dynamical parts are
\begin{equation}
{\cal A}^{(k)}_1=-\frac{1}{2} \epsilon^{ijk} \bar{\psi}_i \gamma_5 \gamma_j, \qquad {\cal A}^{(k)}_2=\frac{1}{2} \epsilon^{ijk}\gamma_5 \gamma_j \psi_i,
\end{equation}
and the explicit form of the symplectic potential is
\begin{equation}
\begin{aligned}
{\cal H}(\psi_0, \bar{\psi}_0, \psi_k, \bar{\psi}_k)&=\frac{1}{2} \epsilon^{ijk} \bar{\psi}_0 \gamma_5 \gamma_i \partial_j \psi_k-\frac{1}{2} \epsilon^{ijk} \bar{\psi}_i \gamma_5 \gamma_0 \partial_j \psi_k-\frac{1}{2} \epsilon^{ijk} \bar{\psi}_i \gamma_5 \gamma_j \partial_k \psi_0 \\
&-\frac{1}{2} \epsilon^{ijk} (\partial_j \bar{\psi}_0) \gamma_5 \gamma_i  \psi_k+\frac{1}{2} \epsilon^{ijk} (\partial_j \bar{\psi}_i) \gamma_5 \gamma_0 \psi_k+\frac{1}{2} \epsilon^{ijk} (\partial_k \bar{\psi}_i) \gamma_5 \gamma_j  \psi_0\\
&-{\rm{i}}  \rm{m} \bar{\psi}_0 \sigma^{0i}\psi_i-{\rm{i}} \rm{m} \bar{\psi}_i \sigma^{i0}\psi_0-{\rm{i}} \rm{m} \bar{\psi}_i \sigma^{ij} \psi_j.
\label{firsthammsslssham12}
	\end{aligned}
	\end{equation}
Like the free massless theory, $\psi_0$ and $\bar{\psi}_0$ are zero modes of the system whose eliminations give rise the constraints 
\begin{equation}
\epsilon^{ijk}(\partial_k \bar{\psi}_i) \gamma_5 \gamma_j-{\rm{i}} \rm{m} \bar{\psi}_i \sigma^{i0} =0, \quad \quad \epsilon^{ijk} \gamma_5 \gamma_i \partial_j \psi_k-{\rm{i}} \rm{m} \sigma^{0i}\psi_i =0.
\label{constrfreemasslessrrs23}
\end{equation}
As was done in the previous section, by decomposing the fields into the local transverse and $\gamma$-traceless parts as in the Eq.(\ref{ttfreemassless12}), the constraints in Eq.(\ref{constrfreemasslessrrs23}) turn into completely transverse ones
\begin{equation}
\epsilon^{ijk}(\partial_k\bar{\psi}^T_i) \gamma_5 \gamma_j-{\rm{i}} \rm{m} \bar{\psi}^T_i \sigma^{i0} =0, \hskip 0.8 cm \epsilon^{ijk} \gamma_5 \gamma_i \partial_j \psi^T_k-{\rm{i}}\rm{m} \sigma^{0i}\psi^T_i=0.
\label{constrfreemasslessrrs32}
\end{equation}
In this case, the longitudinal part reads $\zeta=(\slashed{\partial}+\rm{i} m)\gamma \cdot \psi^T $. Thus, by plugging the Eq.(\ref{ttfreemassless12}) and the transverse constraints Eq.(\ref{constrfreemasslessrrs32}) into the Eq.(\ref{massivelagrangian32}), up to a boundary term, the Lagrangian turns into
\begin{equation}
{\cal L}={\cal A}^{(k)^T}_1 \dot{\psi}^T_k +{\cal A}^{(k)^T}_2 \dot{\bar{\psi}}^T_k +{\rm{i}} \rm{m} \bar{\psi}^T_i \sigma^{i0} \frac{\dot{\zeta}}{\nabla^2}+{\rm{i}} \rm{m} \frac{\dot{\bar{\zeta}}}{\nabla^2} \sigma^{0i} \psi^T_i- {\cal H}^T(\psi^T_k, \bar{\psi}^T_k),
\label{frmassivelagdeclag12}
\end{equation}
where the transverse symplectic coefficients and potential respectively are
\begin{equation}
\begin{aligned}
{\cal A}^{(k)^T}_1&= -\frac{1}{2} \epsilon^{ijk} \bar{\psi}^T_i \gamma_5 \gamma_j, \hskip 1 cm  {\cal A}^{(k)^T}_2=\frac{1}{2} \epsilon^{ijk}\gamma_5 \gamma_j \psi_i,\\
 {\cal H}^T(\psi^T_k, \bar{\psi}^T_k)&= -\frac{1}{2} \epsilon^{ijk} \bar{\psi}^T_i \gamma_5 \gamma_0 \partial_j \psi^T_k+\frac{1}{2}\epsilon^{ijk} \partial_j \bar{\psi}^T_i \gamma_5 \gamma_0 \psi^T_k-{\rm{i}} \rm{m} \bar{\psi}^T_i \sigma^{ij} \psi^T_j.
\end{aligned}
\end{equation}
Observe that the middle two terms in Eq.(\ref{frmassivelagdeclag12}) are not in the symplectic forms. Therefore, by assuming the Darboux transformation
\begin{equation}
\psi^T_k \rightarrow \psi^{'T}_k=e^{2{\rm{i}} \frac{\zeta}{\,\nabla^2} } \psi^T_k, 
\label{darbouxtras}
\end{equation}
with an additional assumption of
\begin{equation}
\epsilon^{ijk} \bar{\psi}^T_i \gamma_5 \gamma_j \psi^T_k=\rm{m} e^{-2{\rm{i}} \frac{\bar{\zeta}}{\,\nabla^2} } \, \bar{\psi}^T_i \sigma^{i0},
\label{adtinlasmpt}
\end{equation}
the undesired terms in Eq.(\ref{frmassivelagdeclag12}) drop and thus we are left with a completely transverse Lagrangian  
\begin{equation}
{\cal L}={\cal A}^{(k)^T}_1 \dot{\psi}^T_k + {\cal A}^{(k)^T}_2 \dot{\bar{\psi}}^T_k-{\cal H}_T(\psi^T_k, \bar{\psi}^T_k)- \lambda_k \phi^k(\psi^T_k, \bar{\psi}^T_k)-\bar{\lambda}_i \bar{\phi}^i(\psi^T_k, \bar{\psi}^T_k).
\label{lagmssvvtrd}
\end{equation} 
Note that the extra condition Eq.(\ref{adtinlasmpt}) is enforced by the Darboux transformation and the constraint equations; otherwise, the coupled terms in the symplectic part could not be decoupled. In fact, it seems there is a lack in the physical interpretation of Eq.(49). Therefore, it will be particularly interesting if one can show that it has a relation with the real constraints or not. Here, as is mentioned in Eq.(\ref{declagmultp}), the remaining variables (i.e., the longitudinal components) are denoted as the Lagrange multipliers
\begin{equation}
\lambda_k=  \frac{\partial_k \zeta}{\nabla^2} \quad , \quad \bar{\lambda}_i= \frac{\partial_i \bar{\zeta}}{\nabla^2},
\end{equation}
such that
\begin{equation}
\begin{aligned}
\phi^k(\psi^T_k, \bar{\psi}^T_k)&= i \epsilon^{ijk} \bar{\psi}^T_i \gamma_5 \gamma_0 \psi^T_j \quad , \quad \bar{\phi}^i(\psi^T_k, \bar{\psi}^T_k)&=-{\rm{i}} \epsilon^{ijk} \bar{\psi}^T_j \gamma_5 \psi^T_k.
\label{extratrmmsslsstryrg}
\end{aligned}
\end{equation}
As noted in \cite{FaddeevJackiw, Jackiwbook}, since the last two terms in the Eq.(\ref{lagmssvvtrd}) cannot be dropped via elimination of constraints anymore, Eq.(\ref{extratrmmsslsstryrg}) corresponds to \emph{the true constraints of the system}. Note also that the true constraints cannot be rewritten as linear combinations of the ones that are obtained during the eliminations of the constraints; otherwise, they would also drop when the eliminations of constraint was performed. These are the constraints that cannot be eliminated anymore. Therefore, setting $\phi^k(\psi^T_k, \bar{\psi}^T_k) $ and $\bar{\phi}^i(\psi^T_k, \bar{\psi}^T_k) $ to zero provides an unconstrained fully traceless Lagrangian 
\begin{equation}
{\cal L}={\cal A}^{(k)^T}_1 \dot{\psi}^T_k + {\cal A}^{(k)^T}_2 \dot{\bar{\psi}}^T_k-{\cal H}_T(\psi^T_k, \bar{\psi}^T_k).
\end{equation}
Thus, with the definition of the dynamical variables $ (\xi_1, \xi_2)=(\psi^T_k,\,\bar{\psi}^T_k) $, the non-vanishing equal-time FJ bracket for the free massive Rarita-Schwinger theory becomes
\begin{equation}
\{\psi^T_i(x),\bar{\psi}^T_k(y) \}_{FJ}=\frac{{\rm{i}}}{2} \gamma_k \gamma_i \gamma_0 \delta^3(x-y).
\end{equation}
which is same as the one found in \cite{Pascalutsa}.

\section{Gauged Massless Rarita-Schwinger Theory}

In this section, we study the massless Rarita-Schwinger field minimally coupled to an external electromagnetic field which is described by the Lagrangian 
\begin{equation}
{\cal L}=-\epsilon^{\lambda\mu\nu\rho} \bar{\psi}_\lambda \gamma_5 \gamma_\mu \overset{\rightarrow}{{\cal D}}_\nu \psi_\rho.
\label{msslsslagfree}
\end{equation}
Here the gauge covariant derivative is $ {\cal D}_\nu =\partial_\nu+g A_\nu$, where $g$ is the relevant coupling constant and $A_\mu$ is an Abelian gauge field. The field equations read
\begin{equation}
\epsilon^{\lambda\mu\nu\rho} \gamma_5 \gamma_\mu \overset{\rightarrow}{{\cal D}}_\nu \psi_\rho =0,\qquad \epsilon^{\lambda\mu\nu\rho} \bar{\psi}_\lambda \overset{\leftarrow}{{\cal D}}_\nu \gamma_5 \gamma_\mu=0. 
\label{fiemasslessgaugedrar12}
\end{equation}
As in the free massless and massive theories, while deducing the some basic properties of the theory, we will only deal with the first of Eq.(\ref{fiemasslessgaugedrar12}). Notice that with the help of the identity in Eq.(\ref{identity}), the Eq.(\ref{fiemasslessgaugedrar12}) turns into
 \begin{equation}
 \slashed{\cal D} \psi^\lambda- {\cal D}^\lambda(\gamma \cdot \psi)-\gamma^\lambda {\cal D} \cdot \psi+\gamma^\lambda \slashed{\cal D} (\gamma \cdot \psi)=0.
 \label{feqmsslssrsgag}
 \end{equation}
Moreover, contracting the Eq.(\ref{feqmsslssrsgag}) with $\gamma_\lambda$ yields
\begin{equation}
\slashed{\cal D} (\gamma \cdot \psi)- {\cal D} \cdot \psi =0.
\label{masslessgagrscosntra12}
\end{equation}
Finally, substituting the Eq.(\ref{masslessgagrscosntra12}) in Eq.(\ref{feqmsslssrsgag}) gives
 \begin{equation}
 \slashed{\cal D} \psi^\lambda- {\cal D}^\lambda(\gamma \cdot \psi)=0.
 \label{diarclikefeqmsslssgaugrs}
 \end{equation}
On the other side, contracting the Eq.(\ref{fiemasslessgaugedrar12}) with ${\cal D}_\lambda$ becomes
\begin{equation}
g \epsilon^{\lambda\mu\nu\rho} \gamma_5 \gamma_\mu F_{\lambda\nu} \psi_\rho =0,
\label{fiemasslessgaugedrar322}
\end{equation}
which is a secondary constraint in the theory and does not provide any further simplification in the field equation in Eq.(\ref{diarclikefeqmsslssgaugrs}).

\subsection*{Symplectic Reduction for Gauged Massless Rarita-Schwinger Theory}

Let us now apply the first-order symplectic formalism to the massless Rarita-Schwinger fields minimally coupled to an external electromagnetic field. For this purpose, let us note that the Lagrangian of the theory in Eq.(\ref{msslsslagfree}) can be recast in a more symmetric form as follows
\begin{equation}
{\cal L}=-\frac{1}{2}\epsilon^{\lambda\mu\nu\rho} \bar{\psi}_\lambda \gamma_5 \gamma_\mu \overset{\rightarrow}{{\cal D}}_\nu \psi_\rho+\frac{1}{2}\epsilon^{\lambda\mu\nu\rho} \bar{\psi}_\lambda \overset{\leftarrow}{{\cal D}}_\nu\gamma_5 \gamma_\mu  \psi_\rho.
\label{gaugedmasslsslagr2}
\end{equation}
Similarly, by splitting the Lagrangian in Eq.(\ref{gaugedmasslsslagr2}) into its space and time components, one gets 
\begin{equation}
{\cal L}={\cal A}^{(k)}_1 \dot{\psi}_k + {\cal A}^{(k)}_2 \dot{\bar{\psi}}_k-{\cal H}(\psi_0, \bar{\psi}_0, \psi_k, \bar{\psi}_k, A_0, A_k),
\label{timspacdecfremsslsssrs124}
\end{equation}
where the symplectic coefficients are
\begin{equation}
{\cal A}^{(k)}_1=-\frac{1}{2}\epsilon^{ijk} \bar{\psi}_i \gamma_5 \gamma_j, \hskip 1 cm {\cal A}^{(k)}_2=\frac{1}{2}\epsilon^{ijk} \gamma_5 \gamma_j \psi_i,
\end{equation}
and the related symplectic potential is 
\begin{equation}
\begin{aligned}
{\cal H}(\psi_0, \bar{\psi}_0, \psi_k, \bar{\psi}_k, A_0, A_k)&=\frac{1}{2}\epsilon^{ijk} \bar{\psi}_0 \gamma_5 \gamma_i \partial_j \psi_k-\frac{1}{2}\epsilon^{ijk} \bar{\psi}_i \gamma_5 \gamma_0 \partial_j \psi_k -\frac{1}{2}\epsilon^{ikj} \bar{\psi}_i \gamma_5 \gamma_k \partial_j \psi_0 \\
&-\frac{1}{2}\epsilon^{ijk} \partial_j \bar{\psi}_0 \gamma_5 \gamma_i \psi_k +\frac{1}{2}\epsilon^{ijk} \partial_j \bar{\psi}_i \gamma_5 \gamma_0 \psi_k+\frac{1}{2}\epsilon^{ikj} \partial_j \bar{\psi}_i \gamma_5 \gamma_k \psi_0 \\
& + g\epsilon^{ijk} \bar{\psi}_i \gamma_5 \gamma_j A_0 \psi_k+g\epsilon^{ijk} \bar{\psi}_0 \gamma_5 \gamma_i A_j \psi_k - g\epsilon^{ijk} \bar{\psi}_i \gamma_5 \gamma_0 A_j \psi_k \\
& -g\epsilon^{ikj} \bar{\psi}_i \gamma_5 \gamma_k A_j \psi_0 .
\end{aligned}
\end{equation} 
Note that although the gauge fields are non-dynamical variables, due to being external potentials, one \emph{cannot} vary and then impose these variations to be vanished. Otherwise, as in the Quantum Electromagnetic Dynamics with external potential, the gauge field current would be enforced to be zero which is not a desired situation. Hence, as in the free theories, here $\psi_0$ and $ \bar{\psi}_0$ are the only zero modes of the theory: Therefore, variations with respect to $ \psi_0 $ and $\bar{\psi}_0$ respectively give the following constraint equations
\begin{equation}
\epsilon^{ikj} \partial_j \bar{\psi}_i \gamma_5 \gamma_k-g\epsilon^{ikj} \bar{\psi}_i \gamma_5 \gamma_k A_j=0, \hskip 0.3 cm \epsilon^{ijk}\gamma_5 \gamma_i \partial_j \psi_k+g\epsilon^{ijk} \gamma_5 \gamma_i A_j \psi_k=0.
\label{constrfreemasslessrrs233} 
\end{equation}
As was done in the free theories, by decomposing the fields into the local transverse and $\gamma$-traceless parts as in Eq.(\ref{ttfreemassless12})\footnote{Notice that, in this case, the longitudinal part becomes $ \zeta = (\slashed{\partial}+g \, \gamma \cdot A ) \gamma \cdot \psi^T-g A \cdot \psi^T $.} and using the constraints in Eq.(\ref{constrfreemasslessrrs233}) as well as by assuming the Darboux transformation (\ref{darbouxtras}), with an additional assumption of
\begin{equation}
{\rm{i}}\epsilon^{ijk} \bar{\psi}^T_i \gamma_5 \gamma_j \psi^T_k= g e^{-2{\rm{i}} \frac{\bar{\zeta}}{\,\nabla^2} } \epsilon^{ijk} \bar{\psi}^T_i \gamma_5 \gamma_k A_j,
\end{equation}
the Lagrangian (\ref{timspacdecfremsslsssrs124}) turns into a completely transverse one 
\begin{equation}
{\cal L}={\cal A}^{(k)^T}_1 \dot{\psi}^T_k + {\cal A}^{(k)^T}_2 \dot{\bar{\psi}}^T_k-{\cal H}_T(\psi^T_k, \bar{\psi}^T_k)-\lambda_k \phi^k(\psi^T_k, \bar{\psi}^T_k) -\bar{\lambda}_i \bar{\phi}^i(\psi^T_k, \bar{\psi}^T_k),
\label{fullttlaggaugrs}
\end{equation}
where the transverse symplectic coefficients and potential read
\begin{equation} 
\begin{aligned}
{\cal A}^{(k)^T}_1&=-\frac{1}{2} \epsilon^{ijk} \bar{\psi}^T_i \gamma_5 \gamma_j, \qquad {\cal A}^{(k)^T}_2=\frac{1}{2} \epsilon^{ijk}\gamma_5 \gamma_j \psi^T_i, \\
{\cal H}_T(\psi^T_k, \bar{\psi}^T_k) &=-\frac{1}{2}\epsilon^{ijk} \bar{\psi}^T_i \gamma_5 \gamma_0 \partial_j \psi^T_k+\frac{1}{2}\epsilon^{ijk} \partial_j \bar{\psi}^T_i \gamma_5 \gamma_0 \psi^T_k + g\epsilon^{ijk} \bar{\psi}^T_i \gamma_5 \gamma_j A_0 \psi^T_k- g\epsilon^{ijk} \bar{\psi}^T_i \gamma_5 \gamma_0 A_j \psi^T_k.
\end{aligned}
\end{equation}
Note that the symplectic potential also contains gauge field parts. Furthermore, as is given in (\ref{declagmultp}), the remaining variables (i.e., the longitudinal components) are denoted as the Lagrange multipliers
\begin{equation}
\lambda_k=  \frac{\partial_k \zeta}{\nabla^2} \quad , \quad \bar{\lambda}_k= \frac{\partial_i \bar{\zeta}}{\nabla^2},
\end{equation}
such that 
\begin{equation}
\begin{aligned}
\phi^k(\psi^T_k, \bar{\psi}^T_k)&= i \epsilon^{ijk} \bar{\psi}^T_i \gamma_5 \gamma_0 \psi^T_j+ g  \epsilon^{ijk} \bar{\psi}^T_i \gamma_5 \gamma_j A_0 +g  \epsilon^{ijk}  \bar{\lambda}_i \gamma_5 \gamma_j A_0-g\epsilon^{ijk} \bar{\psi}^T_i \gamma_5 \gamma_0 A_j\\
\bar{\phi}^i(\psi^T_k, \bar{\psi}^T_k)&=-{\rm{i}} \epsilon^{ijk} \bar{\psi}^T_j \gamma_5 \psi^T_k +g \epsilon^{ijk} \gamma_5 \gamma_j A_0 \psi^T_k
-g \epsilon^{ijk} \gamma_5 \gamma_0 A_j \psi^T_k-g \epsilon^{ijk} \gamma_5 \gamma_0 A_j \lambda_k,
\label{extratrmmsslsstry}
\end{aligned}
\end{equation}
which cannot be dropped via elimination of constraints anymore so, according to \cite{FaddeevJackiw, Jackiwbook}, they are the true constraint of the system. Thus, by setting $\phi^k(\psi^T_k, \bar{\psi}^T_k) $ and $\bar{\phi}^i(\psi^T_k, \bar{\psi}^T_k) $ to zero, one arrives at a completely transverse Lagrangian
\begin{equation}
{\cal L}={\cal A}^{(k)^T}_1 \dot{\psi}^T_k + {\cal A}^{(k)^T}_2 \dot{\bar{\psi}}^T_k-{\cal H}_T(\psi^T_k, \bar{\psi}^T_k). 
\end{equation}
Finally, with the definition of the symplectic dynamical variables $ (\xi_1, \xi_2)=(\psi^T_k,\, \bar{\psi}^T_k) $, one obtains the non-vanishing equal-time FJ basic bracket for the gauged massless Rarita-Schwinger theory as follows
\begin{equation}
\{\psi^T_i(x),\bar{\psi}^T_k(y) \}_{FJ}=\frac{{\rm{i}}}{2} \gamma_k \gamma_i \gamma_0 \delta^3(x-y),
\label{fjbracgaugmasslssrs}
\end{equation}
which is consistent with the Pauli-spin-part of the fundamental bracket obtained in \cite{Adler2015} in which Adler studies the Dirac quantization of the \emph{non-Abelian} gauged Rarita-Schwinger theory via the left-chiral component of the fermionic field. One should notice that such a difference is expected because in \cite{Adler2015}, the corresponding gauge fields are non-Abelian variables; however here the gauge fields are Abelian vector fields. 

\section{Gauged Massive Rarita-Schwinger}

In this section, we study the massive Rarita-Schwinger theory minimally coupled to an external electromagnetic field which is described by the Lagrangian 
\begin{equation}
{\cal L}=-\epsilon^{\lambda\mu\nu\rho} \bar{\psi}_\lambda \gamma_5 \gamma_\mu \overset{\rightarrow}{{\cal D}}_\nu \psi_\rho+{\rm{i}} \rm{m} \bar{\psi}_\lambda \sigma^{\lambda\rho}\psi_\rho , 
\label{laggagmassivers}
\end{equation}
where the gauge-covariant derivative is $ {\cal D}_\nu =\partial_\nu+g A_\nu$. Accordingly, the field equations for the independent anti-commuting fermionic fields are
\begin{equation}
\epsilon^{\lambda\mu\nu\rho} \gamma_5 \gamma_\mu \overset{\rightarrow}{{\cal D}}_\nu \psi_\rho-{\rm{i}} \rm{m} \sigma^{\lambda\rho}\psi_\rho  =0,\qquad \epsilon^{\lambda\mu\nu\rho} \bar{\psi}_\lambda \overset{\leftarrow}{{\cal D}}_\nu \gamma_5 \gamma_\mu +{\rm{i}} \rm{m} \bar{\psi}_\lambda \sigma^{\lambda\rho}=0, 
\label{fiemassivegaugedrar}
\end{equation}
which with the help of the identity in Eq.(\ref{identity}) turns into
\begin{equation}
{\rm{i}} [\slashed{\cal D} \psi^\lambda- {\cal D}^\lambda(\gamma \cdot \psi)-\gamma^\lambda {\cal D} \cdot \psi+\gamma^\lambda \slashed{\cal D} (\gamma \cdot \psi)]-{\rm{i}} \rm{m} \sigma^{\lambda\rho} \psi_\rho=0.
\label{massgauddecpemo}
\end{equation}
Moreover, contraction of the equation in Eq.(\ref{massgauddecpemo}) with $\gamma_\lambda$ gives
\begin{equation}
2 {\rm{i}} (\slashed{\cal D} (\gamma \cdot \psi)- {\cal D} \cdot \psi )+3 \rm{m} \gamma \cdot \psi =0.
\label{massgagrscosntra12}
\end{equation}
And contraction of field equation in Eq.(\ref{fiemassivegaugedrar}) with ${\cal D}_\lambda$ becomes 
\begin{equation}
g \epsilon^{\lambda\mu\nu\rho} \gamma_5 \gamma_\mu F_{\lambda\nu} \psi_\rho+\rm{m} [ (\slashed{\cal D} (\gamma \cdot \psi) -{\cal D} \cdot \psi ] =0, 
\label{fiemassivegaugedrar32}
\end{equation}
which with the additional redefinition 
\begin{equation}
F^d= F^d{_\mu}{^\rho}=\epsilon_\mu{^{\rho\lambda}}{_\nu} F_\lambda{^\nu},
\end{equation}
turns into
\begin{equation}
\rm{m} [\slashed{\cal D} (\gamma \cdot \psi) -{\cal D} \cdot \psi ] -g \gamma_5 \gamma \cdot F^d \cdot \psi =0.
\label{massgagrscosntra22}
\end{equation}
Combining Eq.(\ref{massgagrscosntra12}) and Eq.(\ref{massgagrscosntra22}), one gets the secondary constraint that determines the equation of motion of $\psi^0$ component as follows
\begin{equation}
\gamma \cdot \psi =-\frac{2}{3} \rm{m}^{-2}{\rm{i}} g \gamma_5 \gamma \cdot F^d \cdot \psi.
\label{gagmassrsfineq1}
\end{equation}
Observe that using Eq.(\ref{gagmassrsfineq1}) in Eq.(\ref{massgagrscosntra22}) gives the relation
\begin{equation}
{\cal D} \cdot \psi =-(\slashed{\cal D}-\frac{3{\rm{i}} \rm{m}}{2}) \frac{2}{3} \rm{m}^{-2}{\rm{i}} g \gamma_5 \gamma \cdot F^d \cdot \psi.
\label{gagmassrsfineq2}
\end{equation}
Finally, by plugging Eq.(\ref{gagmassrsfineq1}) and Eq.(\ref{gagmassrsfineq2}) into the field equation in Eq.(\ref{massgauddecpemo}), one obtains
\begin{equation}
({\rm{i}} \slashed{\cal D}-\rm{m}) \psi^\lambda+({\rm{i}} {\cal D}^\lambda+\frac{\rm{m}}{2} \gamma^\lambda) \frac{2}{3} \rm{m}^{-2}{\rm{i}} g \gamma_5 \gamma \cdot F^d \cdot \psi=0,
\label{dirclikfeqmassivegagrs1}
\end{equation}
which is the equation that is used by Velo and Zwanziger in deducing the acausal wave propagation of the solution by finding the future-directed normals to the surfaces at each point \cite{VeloZwanziger}.

\subsection*{Symplectic Reduction for Gauged Massive Rarita-Schwinger Theory}

Finally, let us apply FJ symplectic Hamiltonian reduction to the massive Rarita-Schwinger field minimally coupled to an external electromagnetic field. In order to do so, let us rewrite the Lagrangian in Eq.(\ref{laggagmassivers}) in a more symmetric form: 
\begin{equation}
{\cal L}=-\frac{1}{2}\epsilon^{\lambda\mu\nu\rho} \bar{\psi}_\lambda \gamma_5 \gamma_\mu \overset{\rightarrow}{{\cal D}}_\nu \psi_\rho+\frac{1}{2}\epsilon^{\lambda\mu\nu\rho} \bar{\psi}_\lambda \overset{\leftarrow}{{\cal D}}_\nu\gamma_5 \gamma_\mu  \psi_\rho+{\rm{i}} \rm{m} \bar{\psi}_\lambda \sigma^{\lambda\rho}\psi_\rho.
\label{gaugedmassivelag}
\end{equation}
Subsequently, by splitting Lagrangian in Eq.(\ref{gaugedmassivelag}) into its space and time components, one gets
\begin{equation}
{\cal L}={\cal A}^{(k)}_1 \dot{\psi}_k + {\cal A}^{(k)}_2 \dot{\bar{\psi}}_k-{\cal H}(\psi_0, \bar{\psi}_0, \psi_k, \bar{\psi}_k, A_0, A_k),
\label{timspacdecfremsslsssrs12}
\end{equation}
where the symplectic coefficients are
\begin{equation}
{\cal A}^{(k)}_1 \dot{\psi}_k=-\frac{1}{2}\epsilon^{ijk} \bar{\psi}_i \gamma_5 \gamma_j, \hskip 1 cm {\cal A}^{(k)}_2=\frac{1}{2}\epsilon^{ijk} \gamma_5 \gamma_j \psi_i,
\end{equation}
and the relevant Hamiltonian $ {\cal H}(\psi_0, \bar{\psi}_0, \psi_k, \bar{\psi}_k, A_0, A_k) $ is
\begin{equation}
\begin{aligned}
{\cal H}(\psi_0, \bar{\psi}_0, \psi_k, \bar{\psi}_k, A_0, A_k)&=\frac{1}{2}\epsilon^{ijk} \bar{\psi}_0 \gamma_5 \gamma_i \partial_j \psi_k-\frac{1}{2}\epsilon^{ijk} \bar{\psi}_i \gamma_5 \gamma_0 \partial_j \psi_k -\frac{1}{2}\epsilon^{ikj} \bar{\psi}_i \gamma_5 \gamma_k \partial_j \psi_0 \\
&-\frac{1}{2}\epsilon^{ijk} \partial_j \bar{\psi}_0 \gamma_5 \gamma_i \psi_k  +\frac{1}{2}\epsilon^{ijk} \partial_j \bar{\psi}_i \gamma_5 \gamma_0 \psi_k+\frac{1}{2}\epsilon^{ikj} \partial_j \bar{\psi}_i \gamma_5 \gamma_k \psi_0 \\
&-{\rm{i}} \rm{m} \bar{\psi}_0 \sigma^{0i}\psi_i-{\rm{i}} \rm{m} \bar{\psi}_i \sigma^{i0}\psi_0-{\rm{i}} \rm{m} \bar{\psi}_i \sigma^{ij}\psi_j + g\epsilon^{ijk} \bar{\psi}_i \gamma_5 \gamma_j A_0 \psi_k \\
&+g\epsilon^{ijk} \bar{\psi}_0 \gamma_5 \gamma_i A_j \psi_k- g\epsilon^{ijk} \bar{\psi}_i \gamma_5 \gamma_0 A_j \psi_k-g\epsilon^{ikj} \bar{\psi}_i \gamma_5 \gamma_k A_j \psi_0. 
\end{aligned}
\end{equation} 
Note that as is emphasized in the massless gauged part, since the gauge fields are external potentials, one is not allowed to set their variation to zero. Hence, here $\psi_0, \, \, \bar{\psi}_0$ are the only Lagrange multipliers that induce constraints on the system. Therefore, eliminations of constraint yield 
\begin{equation}
\epsilon^{ikj} \partial_j \bar{\psi}_i \gamma_5 \gamma_k-{\rm{i}} \rm{m} \bar{\psi}_i \sigma^{i0}-g\epsilon^{ikj} \bar{\psi}_i \gamma_5 \gamma_k A_j=0, \hskip 0.3cm \epsilon^{ijk}\gamma_5 \gamma_i \partial_j \psi_k-{\rm{i}} \rm{m} \sigma^{0i}\psi_i+g\epsilon^{ijk} \gamma_5 \gamma_i A_j \psi_k=0.
\label{gaugmassvrascons}
\end{equation}
Like the free massive theory, by decomposing the dynamical components into the local transverse and traceless parts as in Eq.(\ref{ttfreemassless12})\footnote{In this case,  from the constraint equation, one finds 
\begin{equation}
\zeta = (\slashed{\partial}+{\rm{i}} \rm{m} +g \, \gamma \cdot A ) \gamma \cdot \psi^T-g A \cdot \psi^T. 
\end{equation}} as well as using constraints in Eq.(\ref{gaugmassvrascons}) and the Darboux transformation (\ref{darbouxtras}), with an additional assumption of
\begin{equation}
{\rm{i}}\epsilon^{ijk} \bar{\psi}^T_i \gamma_5 \gamma_j \psi^T_k= e^{-2{\rm{i}} \frac{\bar{\zeta}}{\,\nabla^2} }(\rm{i} \rm{m} \bar{\psi}_i \sigma^{i0}+ g \epsilon^{ijk} \bar{\psi}^T_i \gamma_5 \gamma_k A_j),
\end{equation}
the Lagrangian, up to a boundary term, turns into 
\begin{equation}
{\cal L}={\cal A}^{(k)^T}_1 \dot{\psi}^T_k + {\cal A}^{(k)^T}_2 \dot{\bar{\psi}}^T_k-{\cal H}_T(\psi^T_k, \bar{\psi}^T_k)- \lambda_k \phi^k(\psi^T_k, \bar{\psi}^T_k)-\bar{\lambda}_i \bar{\phi}^i(\psi^T_k, \bar{\psi}^T_k) .
\label{onelastmainres}
\end{equation}
Here the transverse symplectic coefficients and potential are
\begin{equation}
\begin{aligned}
{\cal A}^{(k)^T}_1&=-\frac{1}{2} \epsilon^{ijk} \bar{\psi}^T_i \gamma_5 \gamma_j, \qquad {\cal A}^{(k)^T}_2=\frac{1}{2} \epsilon^{ijk}\gamma_5 \gamma_j \psi^T_i ,\\
{\cal H}_T(\psi^T_k, \bar{\psi}^T_k) &=-\frac{1}{2}\epsilon^{ijk} \bar{\psi}^T_i \gamma_5 \gamma_0 \partial_j \psi^T_k+\frac{1}{2}\epsilon^{ijk} \partial_j \bar{\psi}^T_i \gamma_5 \gamma_0 \psi^T_k\\
&+ g\epsilon^{ijk} \bar{\psi}^T_i \gamma_5 \gamma_j A_0 \psi^T_k- g\epsilon^{ijk} \bar{\psi}^T_i \gamma_5 \gamma_0 A_j \psi^T_k-{\rm{i}} \rm{m} \bar{\psi}^T_i \sigma^{ij} \psi^T_j.
\end{aligned}
\end{equation}
Notice that, different from the free cases, the symplectic potential involves mass and gauge potentials. Here in the Eq.(\ref{onelastmainres}), as in the previous sections, the Lagrange multipliers are the Longitudinal parts of the vector-spinor field and the corresponding constraints read  
\begin{equation}
\begin{aligned}
\phi^k(\psi^T_k, \bar{\psi}^T_k)&= i \epsilon^{ijk} \bar{\psi}^T_i \gamma_5 \gamma_0 \psi^T_j+ g  \epsilon^{ijk} \bar{\psi}^T_i \gamma_5 \gamma_j A_0 +g  \epsilon^{ijk}  \bar{\lambda}_i \gamma_5 \gamma_j A_0-g\epsilon^{ijk} \bar{\psi}^T_i \gamma_5 \gamma_0 A_j\\
\bar{\phi}^i(\psi^T_k, \bar{\psi}^T_k)&=-{\rm{i}} \epsilon^{ijk} \bar{\psi}^T_j \gamma_5 \psi^T_k +g \epsilon^{ijk} \gamma_5 \gamma_j A_0 \psi^T_k
-g \epsilon^{ijk} \gamma_5 \gamma_0 A_j \psi^T_k-g \epsilon^{ijk} \gamma_5 \gamma_0 A_j \lambda_k,
\label{extratrmmsslsstrymn}
\end{aligned}
\end{equation}
which are same as Eq.(\ref{extratrmmsslsstry}). Similarly, by setting Eq.(\ref{extratrmmsslsstrymn}) to zero \cite{FaddeevJackiw, Jackiwbook}, one arrives at a completely transverse Lagrangian 
\begin{equation}
{\cal L}={\cal A}^{(k)^T}_1 \dot{\psi}^T_k + {\cal A}^{(k)^T}_2 \dot{\bar{\psi}}^T_k-{\cal H}_T(\psi^T_k, \bar{\psi}^T_k),
\end{equation}
whose symplectic part is same as the ones found so far. Thus, with the definition of the dynamical variables $ (\xi_1, \xi_2)=(\psi^T_k,\,\bar{\psi}^T_k) $, the non-vanishing equal-time bracket for the gauged massive Rarita-Schwinger theory becomes
\begin{equation}
\{\psi^T_i(x),\bar{\psi}^T_k(y) \}_{FJ}=\frac{{\rm{i}}}{2} \gamma_k \gamma_i \gamma_0 \delta^3(x-y),
\end{equation}
which is identical to the one found in \cite{Inoue}.

\section{Conclusions}

In this work, we studied $3+1$-dimensional free and Abelian gauged Grassmannian Rarita-Schwinger theories for their massless and massive extensions in the context of Faddeev-Jackiw first-order symplectic formalism. We have obtained the fundamental brackets of theories which are consistent with the some results that we found in the literature but obtained in a more simpler way. The brackets are independent of whether the theories contain mass or gauge field or not, and thus the structure of constraints and symplectic potentials determine characteristic behaviors of the theories. It will be particularly interesting to find proper transformations that will relate the constraints obtained via the Faddeev-Jackiw symplectic method with the ones that are obtained via Dirac method. But since the constraints obtained in both methods are rather complicated, in this paper, we restrict ourselves only to the Faddeev-Jackiw analysis of Rarita-Schwinger theories and leave this as a future work. With the comparison with the literature, we concluded that the Faddeev-Jackiw symplectic approach provides a more economical way in deriving the fundamental brackets for the Rarita-Schwinger theories. In addition to these, we notice that, in contrast to the massive theory, the Dirac field equations for free massless Rarita-Schwinger theory cannot be covariantly deduced.

\section{\label{ackno} Acknowledgments}

We would like to thank Roman Jackiw for suggesting the problem and several useful discussions. We would also like to thank Bayram Tekin for useful suggestions and Markus Schulze, Gilly Elor and Ibrahim Burak Ilhan for critical readings of the paper. S.D. is supported by the TUBITAK 2219 Scholarship.

\section{Appendix A: Transverse and Traceless Decomposition of Fields} 

In this section, let us give the derivations of (\ref{ttdecfrersmsslssd1}) and (\ref{constrfreemasslessrrs22}): To solve the constraint equations, one can decompose the independent fields into its \emph{local transverse and $\gamma$-traceless parts} as
\begin{equation}
\psi_i =\psi^T_i+\hat{\psi}_i \qquad \bar{\psi}_i=\bar{\psi}^T_i+\hat{\bar{\psi}}_i,
\label{ttfreemassless122}
\end{equation}	
where "$T$" and "$\textasciicircum$" stand for the transverse and traceless parts, respectively. Here the $\gamma$-traceless parts are
\begin{equation}
\hat{\psi}_i=\psi_i-\frac{1}{3} \gamma_i \gamma^j \psi_j, \hskip 1 cm \hat{\bar{\psi}}_i=\bar{\psi}_i-\frac{1}{3} \bar{\psi}_j \gamma^j \gamma_i.
\label{ttdecomp}
\end{equation}	
Therefore, we have 
\begin{equation}
\partial^i \psi^T_i=\partial^i \bar{\psi}^T_i=0 \quad \mbox{and} \quad \gamma^i \hat{\psi}_i=\gamma^i \hat{\bar{\psi}}_i=0 .
\label{gammttcond}
\end{equation}
To find how the constraint equations in Eq.(\ref{constrfreemasslessrrs12}) decomposes under Eq.(\ref{ttfreemassless122}), let us focus on the following constraint equation  
\begin{equation}
\epsilon^{ijk} \gamma_5 \gamma_i \partial_j \psi_k=0.
	\label{constrfirstap}
\end{equation}
Note that with the identity $\epsilon^{ijk} \gamma_5 \gamma_k=-\gamma^0 \sigma^{ij} $ and Eq.(\ref{ttfreemassless122}), the Eq.(\ref{constrfirstap}) turns into
\begin{equation}
\frac{i\gamma^0}{2} [\gamma^k, \gamma^j] \partial_j  (\psi^T_k+\hat{\psi}_k) =0. 
\end{equation}
Furthermore, by using the relation
\begin{equation}
 [\gamma^k, \gamma^j]= \{\gamma^k, \gamma^j \}-2\gamma^j \gamma^k=2 (\eta^{kj}-\gamma^j \gamma^k ),
\end{equation}
and the transverse and traceless properties of the fields in Eq.(\ref{gammttcond}), one gets
\begin{equation}
i\gamma^0\Big(\partial^k \hat{\psi}_k-\gamma^j\gamma^k \partial_j \psi^T_k\Big) =0.
\end{equation}
Notice that after contraction with $ i \gamma_0 $ and relabeling of the dummy indices, it becomes
\begin{equation}
\partial^i \hat{\psi}_i-\gamma^m \gamma^n \partial_m  \psi^T_n =0,
\end{equation}
which yields
\begin{equation}
\hat{\psi}_i=\frac{\partial_i \zeta}{\nabla^2} \quad \mbox{where} \quad \zeta=\gamma^m \gamma^n \partial_m  \psi^T_n.
\end{equation}
This structure is also valid for the other theories. The only difference arises in the definition of $\zeta$ which we give its explicit form in each section. Finally, by substituting this result into the constraint in Eq.(\ref{constrfirstap}), it turns into
\begin{equation}
		\epsilon^{ijk} \gamma_5 \gamma_i \partial_j \psi^T_k+\epsilon^{ijk} \gamma_5 \gamma_i \frac{\partial_j \partial_k \zeta}{\nabla^2}=0. 
\end{equation}
Because of the symmetric and anti-symmetric contraction in "$ j, k $" indices, the second term drops, and we are left with the transverse constraint equation
\begin{equation}
	\epsilon^{ijk} \gamma_5 \gamma_i \partial_j \psi^T_k=0.
\end{equation}	

\section{Appendix B: Faddeev-Jackiw Hamiltonian Reduction for Constrained and Unconstrained Systems} 

In this section, we review the Faddeev-Jackiw symplectic first-order formalism which was introduced particularly to quantize the constrained systems \cite{FaddeevJackiw, Jackiwbook}. The method works on the first-order Lagrangian and does not require any classification of constraints. To better understand how the method works, let us consider
\begin{equation}
{\cal L}=p_\alpha \dot{q}^\alpha-H(p,q), \qquad \alpha =1,\dots n. 
\label{fjmlgr}
\end{equation} 
With the definition of $2n$-component phase-space coordinates
\begin{equation}
\xi^\alpha=p_\alpha, \quad \alpha =1, \cdots, n \quad \mbox{and} \quad \xi^\beta =q^\beta, \quad \beta =n+1, \cdots, 2n,
\end{equation}
Eq.(\ref{fjmlgr}) can be rewritten as a Lagrangian one-form 
\begin{equation}
{\cal L}dt=\frac{1}{2} \xi^\alpha f^0_{\alpha \beta} d\xi^\beta -V(\xi) dt.
\label{oneformlagr}
\end{equation} 
Here the symplectic $2n \times 2n$ matrix is 
\[ f^0_{\alpha \beta}= \left( \begin{array}{cc}
\,\, \, 0  &\,\, \mbox{I}  \\
-\mbox{I} & \,\, 0 \end{array} \right)_{\alpha \beta }, \]
where $\mbox{I}$ is the identity matrix; $A_0 \equiv \frac{1}{2} \xi^\alpha f^0_{\alpha \beta} d\xi^\beta $ is the \emph{canonical one-form}; $f^0 \equiv d A_0 \equiv \frac{1}{2} f^0_{\alpha\beta} d\xi^\alpha d\xi^\beta $ is the \emph{symplectic two-form}. Note that $f^0$ is constant \cite{FaddeevJackiw, Jackiwbook}. But, in general, the symplectic two-form does not have to be constant. Therefore, let us now consider the following generic Lagrangian 
\begin{equation}
{\cal L} dt= A_\alpha d\xi^\alpha-H(\xi) dt, \qquad \alpha=1, \cdots, 2n,
\label{generoneformlag} 
\end{equation}
where $A_\alpha $ is an arbitrary one-form. The variation of Eq.(\ref{generoneformlag}) with respect to $\xi$ yields
\begin{equation}
f_{\beta\alpha} \dot{\xi}_\alpha =\frac{\partial H}{\partial \xi_\beta} \hskip 0.5cm \mbox{where} \hskip 0.5 cm f_{\beta \alpha}=\frac{\partial A_\alpha}{\partial \xi_\beta}-\frac{\partial A_\beta}{\partial \xi^\alpha}.
\label{fjemos}
\end{equation}
In the case of when the symplectic matrix is nonsingular, Eq.(\ref{fjemos}) becomes
\begin{equation}
\dot{\xi}_\alpha = f^{-1}_{\alpha\beta} \, \frac{\partial H(\xi)}{\partial \xi_\beta}.  
\label{bosonicemo}
\end{equation}
Thus, by using Eq.(\ref{bosonicemo}) and the Poisson brackets for the bosonic variables, one obtains the FJ fundamental brackets as follows
 \begin{equation}
 \{\xi_\beta, \xi_\alpha\}_{FJ}=f^{-1}_{\alpha\beta}. 
 \end{equation}
 Note that, in the case of the Grassmannian variables, using the anti-commutation property of the variables as well as the Poisson brackets for the Grassmannian variables \cite{Casalbuoni}, one has
 \begin{equation}
\dot{\xi}_\alpha =\frac{\partial H(\xi)}{\partial \xi_\beta} (f^{-1})_{\alpha\beta}, 
\end{equation}
and the corresponding fundamental brackets become 
\begin{equation}
\{\xi_\beta, \xi_\alpha\}_{FJ}=-(f^{-1})_{\alpha\beta} .
\label{grassfjquant}
\end{equation}

On the other side, when there are constraints in the system which are induced by the existence of the zero-modes, then the symplectic matrix cannot be inverted. In that case, according to the Darboux's theorem which states that for any given one-form $A=A_\alpha d \xi^\alpha $ where $ \alpha =1, \cdots, N  $, one can always do the following changes in the variables
\begin{equation}
\xi^\alpha \rightarrow (p^\beta, q^\gamma z^\rho), \quad \beta,\, \gamma =1, \cdots, n, \quad \rho =1, \cdots, N-2n, 
\label{diag}
\end{equation}
so that $A$ turns into $A=A_\alpha dq^\alpha $.  As is seen above, when there is no constraint, Eq.(\ref{diag}) diagonalizes $f_{\alpha\beta}.$ However when there are constraints, only a $ 2n \times 2n $ sub-block of $ f_{\alpha\beta} $ \emph{diagonalizes} and the remaining $N-2n$ degrees of freedom (corresponding the zero-modes $z^\rho $) will not be in the symplectic form \cite{FaddeevJackiw, Jackiwbook}; yet they occur in the rest of the Lagrangian:
\begin{equation}
{\cal L}=p_\alpha dq^\alpha -\Phi(p,q,z) dt.
\end{equation}
The equations $\frac{\partial \Phi}{\partial z^\alpha}=0$ can be used to eliminate the zero-modes of $z$'s only if $\frac{\partial^2 \Phi}{\partial z^\rho \partial z^\beta}$ is nonsingular. In the generic case, after diagonalization and elimination of $z$'s as many as possible, one ultimately arrives at
\begin{equation}
{\cal L}=p_\alpha \dot{q}^\alpha-H(p,q)-\lambda_\rho \phi^\rho (p,q),
\label{declagmultp}
\end{equation}
where the remaining $z$'s are denoted by $ \lambda_\rho $ (namely, Lagrange multipliers) and the $\phi^\rho$ are \emph{the only true constraints in the system}
\begin{equation}  
\phi^\rho=0.
\end{equation}

\subsection{Symplectic Reduction for Dirac Theory of spin-$\frac{1}{2}$ fields}

In this section, to see how the method works, we provide FJ Hamiltonian reduction for the Dirac theory for the spin-$\frac{1}{2}$ theory as an example. For this purpose, let us note that the Lagrangian can be written: 
\begin{equation}
{\cal L}=-\frac{i}{2} \bar{\psi}\overset{\leftarrow}{\slashed{\partial}}\psi+\frac{i}{2}\bar{\psi}  \overset{\rightarrow}{\slashed{\partial}} \psi -m \bar{\psi} \psi . 
\label{diraclag}
\end{equation}
As mentioned above, we assume that the independent dynamical variables are anti-commuting Grassmann variables. In order to pass to the symplectic analysis of the system, one needs to separate the dynamical components from the non-physical ones by splitting the Lagrangian (\ref{diraclag}) into its time and space components. In doing so, one arrives at
\begin{equation}
{\cal L}=\frac{i}{2} \gamma^0 \psi \dot{\bar{\psi}}+\frac{i}{2} \gamma^0 \bar{\psi} \dot{\psi}-\Big[\frac{i}{2}\partial_i\bar{\psi} \gamma^i  \psi-\frac{i}{2}\bar{\psi} \gamma^i \partial_i \psi+m \bar{\psi} \psi \Big],
\end{equation}
whose variation, up to a boundary term, yields
\begin{equation}
\delta {\cal L}= \delta \psi \Big (i \gamma^0 \dot{\bar{\psi}} \Big)+\delta \bar{\psi} \Big(i \gamma^0  \dot{\psi}\Big)- \Big[ \delta \bar{\psi} (-i \gamma^i \partial_i \psi +m \psi) +\delta \psi (-i\gamma^i\partial_i \bar{\psi}-m\bar{\psi} )\Big],
\label{diracsymp}
\end{equation}
from which one gets the Dirac field equations as follows
\begin{equation}
i \gamma^0 \dot{\bar{\psi}}=-i\gamma^i\partial_i \bar{\psi}-m\bar{\psi} , \qquad i \gamma^0  \dot{\psi}=-i \gamma^i \partial_i \psi +m \psi.
\label{diracfeqfj}
\end{equation}
As is seen from Eq.(\ref{diracsymp}) and Eq.(\ref{diracfeqfj}), the symplectic matrix for the Dirac theory and its inverse are
\[ f_{\alpha\beta }= \left( \begin{array}{cc}
0  & i \gamma^0 \\
i \gamma^0  & 0 \end{array} \right), \quad  f^{-1}_{\alpha\beta }= \left( \begin{array}{cc}
0  &- i \gamma^0 \\
-i \gamma^0  & 0 \end{array} \right)=- f_{\alpha\beta }. \]
One should observe that, in contrast to the bosonic case, the symplectic matrix for the Grassmannian variables is symmetric and the fundamental brackets are defined as follows
\begin{equation}
\{\xi_\beta, \xi_\alpha\}_{FJ}=-(f^{-1})_{\alpha\beta},
\end{equation}
from which one gets the basic bracket for the Dirac theory 
\begin{equation}
\{\psi, \bar{\psi}\}_{FJ}=i\gamma^0.
\end{equation}
This is also valid for the massless theory. Note that since the theory does not have any gauge redundancy, one does not need to assume any gauge-fixing.

		\end{document}